\documentclass[12 pt]{article}
\usepackage{graphicx}
\pdfoutput=1
\usepackage{amsmath}
\usepackage[ruled,vlined]{algorithm2e}

\usepackage{dcolumn}
\usepackage{amsmath} 
\usepackage{amssymb} 
\usepackage{enumerate}
\usepackage{amsthm}
\usepackage{float} 
\usepackage{capt-of} 
\usepackage{sidecap} 
\sidecaptionvpos{figure}{c} 
\usepackage{caption} 
\usepackage{commath} 
\usepackage{cancel} 
\usepackage{anysize} 
\usepackage{appendix} 
\usepackage{tocbibind}
\usepackage{physics} 
\usepackage{multirow}
\usepackage{glossaries}

\newtheorem{theorem}{Theorem}

\newtheorem{lemma}{Lemma}
\newtheorem{proposition}{Proposition}

\begin{document}
	
\title{Synchronization in a class of chaotic systems}

\author{	J. Telenchana and A. Acosta\\
\small		School of Mathematical and Computational Sciences, \\
\small		Yachay Tech University, Ecuador.\\ \\
        	P. Garc\'{\i}a\\
\small		Laboratorio de Sistemas Complejos,
\small		Departamento de F\'{\i}sica Aplicada,\\
\small		Facultad de Ingenier\'{\i}a, 
\small		Universidad Central de Venezuela.\\	
\small		Red Iberoamericana de Investigadores en
\small		Matemáticas Aplicadas a  Datos. AUIP.
}

\date{}
	
\maketitle
\thispagestyle{empty}

\begin{abstract}
	In this work, the synchronization problem of a master-slave system of autonomous ordinary differential equations (ODEs) is considered. Here, the systems are, chaotic with a nonlinearity represented by a piecewise linear function, non-identical and linearly coupled.  
	
	The idea behind our methodology is quite simple: we couple the systems with a linear function of the difference between the states of the systems and we propose a formal solution for the ODE that governs the evolution of that difference and then we determine what the parameters should be
	of the coupling function, so that the solution of that ODE is a fixed point close to zero.
	
	As the main result, we obtain conditions for the coupling function that guarantize the synchronization, based on a suitable descomposition of the system joined to a fixed point theorem. 
	
	The scheme seems to be valid for a wide class of chaotic systems of great practical utility.
\end{abstract}

\section{\label{sec:level1}INTRODUCTION}

Synchronization in dynamical systems describes the correlated evolution of two or more systems under specific conditions for a time interval \cite{Luo2009}. We adopt this, as the definition of synchronization although, as suggested by  \cite{Brown2000}, it is very hard to establish a single definition of synchronization that encompasses each and every known example of synchronization. 
This phenomenon has been studied since the century 17th, when Christiaan Huygens gave a detailed description about the synchronization of dynamical systems formed by two pendulum-clocks hanging from simple support, see for example \cite{Willms}. 

Over time, the interest in the study of synchronization has increased. This is due to its relevant manifestation in fields such as Communications\cite{Lars}, Nanotechnology\cite{Waried} or Biology\cite{Perez1,Perez2}. 

The synchronization, in general, can occur spontaneously between interconnected parts of a particular system as in \cite{Willms}, or it can be induced, establishing conections in the systems that promote it, both on localized \cite{Pecora-review, Acosta2001} or espatially extended  systems \cite{Acosta2013, Garcia, DeAbreu}. In \cite{Perez1} and \cite{Perez2}, there are two very interesting examples from human physiology.

When considering chaotic dynamical systems that are uncoupled, sensitivity to initial conditions prevents spontaneous synchronization. Thus, a problem of synchronization of chaotic systems can be posed as the choice of coupling for the systems and the subsequent conditions to obtain the desired correlations as time evolves.

In this direction, the first work that made a very important impact corresponds to Fujisaka and Yamada \cite{Fujisaka}, although the best known reference is by Pecora and Carrol  \cite{Pecora1990}. Since then, the number of works describing fundamental aspects of the phenomenon or its applications has been increasing significantly, see for example the reviews \cite{Pecora, Boccaleti} and the references that are found there that add up to several hundred.

In the particular case of synchronization of piece-wise linear chaotic systems, there are many synchronization strategies with a significant number of them based on control strategies or using the Lyapunov approach, see for example \cite{Almeida, Mkaouar} and the references found there.

Here, we deal with the problem of establishing conditions that guarantee synchronization between two different chaotic systems. Our scenario, consider a master-slave nonidentical chaotic systems where the nonlinearities are represented by piecewise linear functions and we use as a example the Chua´s equations \cite{Chua1992}.
These systems are simple electronic circuit that exhibits classic chaotic behavior and we will use them as representatives of a class of chaotic systems with nonlinearity given by piecewise linear functions \cite{Yang, Brown, LAKSHMANAN}

The solution to the synchronization problem mentioned above, in this case, begins with the selection of a linear coupling, although we believe that non-linear couplings\cite{Feketa} can be included in this same scheme, function. For this kind of coupling, several strategies have been proposed to achieve synchronization, see for instance \cite{Ogorzalek}.

Once the form of the coupling function has been chosen, there are sufficient conditions that the parameters of that function must satisfy, so that synchronization is achieved.

In order to expose our synchronization problem and its solution, we have organized the work as follows:

In section 2, we present the problem to be considered. Specifically the type of dynamical systems and the type of coupling function between them. 
Section 3 is devoted to develop the syncronization strategy. In section 4, we present the application of the results in Chua's equations as a representative of a class of piece-wise linear nonlinearity. 
In section 4,  a numerical implementation of the before mentioned results is shown.	
In section 5, we present some final remarks. There we highlight the characteristics that the systems considered in our work have, and also mention considerations that may lead us to carry out further research.
Finally, with the intention of making the exposition more continuous, in the appendix, we offer the proofs for all the propositions, lemmas and the theorem associated with our main result, in cases where it is necessary.

\section{The synchronization problem}

Let us consider as synchronization problem, the selection of the coupling parameter $ \nu $  in the following  master-slave system:
\begin{align}
\mathbf{\Dot{x}}&= {\bf f}(\mathbf{x}, \Bar{\mu}), \label{2master}
\\ \mathbf{\Dot{y}}&= {\bf f}(\mathbf{y}, \mu)+\nu(\mathbf{y}-\mathbf{x}),  \label{2slave}  
\end{align}

\noindent
where $\nu$ is a real constant, $ \Bar{\mu}, \mu $ are vector parameters in $\mathbb{R}^m$ and $ {\bf f}: \mathbb{R}^n\times\mathbb{R}^m\rightarrow\mathbb{R}^n$ is a continuous function. 

To pose the problem from which we will be able to guarantee conditions that imply the synchronization of the system (\ref{2master})-(\ref{2slave}), we will begin by considering a bounded solution  $ {\bf x}(t, {\bf x}_0 , \bar{\mu}) $ of (\ref{2master}). Here  $ {\bf x}(t, {\bf x}_0, \bar{\mu}) $  stands by a solution such that at $t=0 $ gives  ${\bf x}(0, {\bf x}_0, \bar{\mu})= {\bf x}_0 $. Now, consider the following  transformation

\begin{equation} \label{Transformation}
\mathbf{z}= {\bf y}-{\bf x}(t, {\bf x}_0, \bar{\mu}).
\end{equation}

If we consider $\mathbf{y}$ as a slave solution, i.e., solution of $(\ref{2slave})$ with input ${\bf x}(t, {\bf x}_0, \bar{\mu})$, then the previous transformation yields the non-autonomous equation 

\begin{align}\label{zpuntito}
\mathbf{\dot{z}}&=\nu \mathbf{z}+{\bf f}( \mathbf{z} + {\bf x}(t, {\bf x}_0, \bar{\mu}), \mu)-{\bf f}( {\bf x}(t, {\bf x}_0, \bar{\mu}),\bar{\mu}  )\notag
\\&:=  F \left( t , {\bf{z}} , \nu ,  \mu , \bar{\mu} \right) .
\end{align}

We will now focus on the equation 

\begin{equation} \label{NonAut}
{\bf{\dot{z}}}  = F \left( t , {\bf{z}} , \nu , \mu , \bar{\mu} \right),
\end{equation}

\noindent
where  $F :\left[ 0 , \infty \right) \times {\mathbb{R}}^{n} \times {\mathbb{R}} \times {\mathbb{R}}^{m} \times {\mathbb{R}}^{m} \rightarrow {\mathbb{R}}^{n} $ is a continuous function. It is assumed that $F$ can be decomposed as the sum of three functions  

\begin{equation} \label{Decomposition}
F \left( t , {\bf{z}} , \nu , \mu , \bar{\mu} \right)= B {\bf{z}}+ G \left( t , \mu , \bar{\mu} \right) + H \left( t , {\bf{z}}, \mu \right).
\end{equation}

In what follows, we will choose the conditions that $F$ must meet and that limits our result to that class of functions and couplings. For the previous decomposition, we establish the following hypotheses  ($H_i$):

\begin{itemize}
	\item [$H_1$.] $B$ is a constant real  $n \times n$ matrix for which all the eigenvalues have negative real part. 
	
	\item [$H_2$.] $ G: \left[ 0 , \infty \right) \times  {\mathbb{R}}^{m} \times {\mathbb{R}}^{m} \rightarrow {\mathbb{R}}^{n}$ is  continuous and satisfies: if given $ \varepsilon >0 $, then $ \delta >0 $ exists such that 
	\begin{equation*}
	\norm{G \left( t , \mu , \bar{\mu} \right)} < \varepsilon,
	\end{equation*}
	
	for any $t \geq 0$ and $ \mu , \bar{\mu}$ with $ \norm{\mu - \bar{\mu}} < \delta$.
	
	\item [$H_3$.] $ H : \left[ 0 , \infty \right) \times  {\mathbb{R}}^{n} \times {\mathbb{R}}^{m} \rightarrow {\mathbb{R}}^{n}$ is a continuous function such that $H \left( t , 0, \mu \right)=0$ for any $ \mu \in {\mathbb{R}}^{m} $ and $t \geq 0$. Also it satisfies the following type of Lipschitz condition: for any $ \mu \in {\mathbb{R}}^{m} $ there is a positive constant $L = L \left(\mu  \right)$ such that
	\begin{equation*}
	\norm{H \left( t , {\bf{z}}_1, \mu \right) - H \left( t , {\bf{z}}_2 , \mu \right)} \leq L \norm{{\bf{z}}_1 - {\bf{z}}_2 } ~,\quad t \geq 0.
	\end{equation*}
\end{itemize}

These assumptions establish the characteristics of $F$ and condition our results. Regarding $ H_1 $,  the first proposition in this work shows an interesting result that will be useful later.
\begin{proposition}
	If $B$ is a constant real  $n \times n$ matrix for which all the eigenvalues have negative real part, then there are positive constants $ K , \gamma$ such that 
	\begin{equation} \label{useful}
	\norm{ e^{Bt}{\bf{z}} } \leq Ke^{- \gamma t} \norm{ {\bf{z}} } ,\quad t \geq 0 , \quad{\bf{z}}  \in {\mathbb{R}}^{n} .
	\end{equation}
	\label{Prop1}
\end{proposition}

Hypothesis $H_2$ tells us that the norm of $G$, depending on how close the parameters $\mu$ and $\bar{\mu}$ are, can be made sufficiently small. $H_3$ is a type of condition that is often considered when looking for existence and uniqueness of solutions for differential equations. 

Now, in our problem we study the system $(\ref{NonAut})$ with the decomposition given in $(\ref{Decomposition})$ and under the hypotheses $H_1$, $H_2$ and $H_3$. We pursuit to find solutions of $(\ref{NonAut})$ that, when associated with the transformation $(\ref{Transformation})$, guarantee synchronization of the master-slave system (\ref{2master})$-$(\ref{2slave}). 

\section{The synchronization strategy}
The idea behind our approach begin by proposing a formal solution to system (\ref{NonAut}), in the form of our first lemma here, to later prove that under certain conditions, this solution converges to a fixed point such that the orbit of the slave system is close to the orbit of master system.

\begin{lemma}
	The initial value problem
	\begin{equation} \label{NonAutIVP}
	\left\{\begin{array}{ll}
	{\bf{\dot{z}}}  = F \left( t , {\bf{z}} , \mu , \bar{\mu} \right) \\ {\bf{z}}(0)={\bf{z}}_0,
	\end{array}\right.
	\end{equation}
	
	where $F$ is decomposed as in (\ref{Decomposition}),
	is  equivalent to
	
	\begin{equation}
	\label{EquiIntEq}
	{\bf z}(t) = e^{B t}{\bf{z}}_0 +
	\int_{0}^{t}e^{B(t-s)}N(s , {\bf{z}}(s), \mu , \bar \mu) ds.
	\end{equation}
	\label{Lemm1}
\end{lemma}

\noindent
where, $N(s , {\bf{z}}(s), \mu , \bar \mu) = G(s , \mu , \bar{\mu})+ H(s , {\bf{z}}(s), \mu )$

Suppose $K , \gamma$ are the constants appearing in (\ref{useful}). Let $ \rho >0$ and $ \mu , \bar{\mu} \in  {\mathbb{R}}^{m}$ such that
\begin{equation} \label{Condition2}
\norm{ G \left( t , \mu , \bar{\mu} \right)} < \frac{\rho \gamma}{4K}~, \quad t \geq 0 .
\end{equation}

With this choice of $\rho, \mu$ and $\bar{\mu}$, and for $ {\bf{z}}_{0} \in {\mathbb{R}}^{n}$ satisfying $\displaystyle\norm{{\bf{z}}_{0} } < \frac{\rho}{2K} $  we define

	\begin{equation}
	\mathcal{G}\left(\mathbf{z}_{0}, \rho, \mu, \bar{\mu}\right)  :=  \left\{ \mathbf{z} \in \mathcal{C}_{b}\left([0, \infty), \mathbb{R}^{\mathbf{n}}\right):   
	{\norm{\mathbf{z}}}_{\infty} := \sup _{\mathbf{t} \geq 0} \Vert \mathbf{z}(t)  \Vert \leq \rho \right \} \nonumber
	\end{equation}

\noindent
with $\mathbf{z}(0)=\mathbf{z}_{0}$.

\begin{proposition}
	$\mathcal{G}\left(\mathbf{z}_{0}, \rho, \mu, \bar{\mu}\right)$ is a closed subset of the Banach space $X=\mathcal{C}_{b}\left([0, \infty), \mathbb{R}^{\mathbf{n}}\right)$, bounded continuous functions from $\left[ 0 , \infty \right)$ to ${\mathbb{R}}^{n} $, with the supremum norm.
\end{proposition}

Inspired in Lemma \ref{Lemm1}, for any ${\mathbf{z}} \in  \cal G \left(  \mathbf{z}_{0} , \rho , \mu , \bar{\mu}  \right)$ we define an operator $T{\bf{z}}$ by

\begin{equation}
\left( T {\bf{z}}  \right)(t) =  e^{B t} {\mathbf{z}}_{0} + \int_{0}^{t}e^{B(t-s)} N(s , {\bf{z}}(s), \mu , \bar \mu)) ds,
\end{equation}

\noindent
with $t \geq 0 $.

We have, due to the fact that $ e^{B (\cdot)}{\mathbf{z}_0}$ is continuous and the hypothesis $H_2$, $H_3$,  that $T{\bf{z}}$ is a continuous function for $t \geq 0$. 

Now, we state our main theoretical result. It establishes a relationship, although not explicitly given, between the parameters of the coupling function and the quality of synchronization, defined in this case as the mean square error, between the  states of  the master and slave systems. 

\begin{theorem}
	If
	\begin{equation} \label{otraC}
	\frac{KL}{\gamma} \leq \frac{1}{4},
	\end{equation}
	where $L=L(\bar{\mu})$ is the constant given in $H_3$, then 
	$T$ acts from $\cal G \left( \bf{z}_{0} , \rho , \mu , \bar{\mu}  \right)$ into itself and also has a unique fixed point in $\cal G \left( \bf{z}_{0} , \rho , \mu , \bar{\mu}  \right)$.
\end{theorem}

Next, in order to broaden the perpective of our resuts and highlight more about the scope of those, we apply the Gronwall's lemma. Let us recall Gronwall's lemma:

\begin{lemma}
	Let $M$ be a non-negative constant and let $f$ and $g$ be continuous non-negative functions, for $a \leq t \leq b$, satisfying
	\begin{equation*}
	f(t) \leq M+\int_{a}^{t} f(s)g(s)ds,~ a \leqslant t \leqslant b,
	\end{equation*}
	then
	\begin{equation*}
	f(t) \leq M e^{\int_{a}^{b}g(s)ds} \quad a \leqslant t \leqslant b ~.
	\end{equation*}
	\label{Lemm2}
\end{lemma}


If the operator $T$ is considered as a function of ${\bf{z}}_0$; that is $T=T_{{\bf{z}}_0}$, and ${\bf{{{z}}^{ \star}}}(\cdot , {\bf{z}}_0 )$ is the unique fixed point, then we have that  ${\bf{z^{\star}}}( t , {\bf{z}}_0 )$ is continuous in ${\bf{z}}_0$ uniformly with respect to $t$. In fact: Consider the sets  $\cal G \left( {\bf{z}_0} , \rho , \mu , \bar{\mu}  \right)$,  $\cal G \left( \bf{\tilde{z}_0} , \rho , \mu , \bar{\mu}  \right)$ and  us denote by  $ {\bf{z}}^{\star}:= \bf{{{z}}^{ \star}}(\cdot , {\bf{{z}}_0} )$ and  $ \bf{{\tilde{z}}^{\star}}:=\bf{{\tilde{z}}^{ \star}}(\cdot , {\bf{\tilde{z}}_0} )$ the fixed points of the operator $T$ on $\cal G \left( {\bf{z}_0} , \rho , \mu , \bar{\mu}  \right)$ and  $\cal G \left( {\bf{\tilde{z}}_0} , \rho , \mu , \bar{\mu}  \right)$, respectively. Defining $\Delta {\bf z} = { \bf{{{z}}^{ \star}}}( t , {\bf{{z}}_0} ) - {\bf{{\tilde{z}}^{ \star}}}( t , {\bf{\tilde{z}}_0} )$ we have

\begin{equation}
\Delta {\bf z} =   e^{Bt}({\bf{{z}}_0}-{\bf{\tilde{z}}_0} ) + 
\int_{0}^{t} e^{B(t-s)}\Delta H^*  ~ds. 
\end{equation}

\noindent
with, $\Delta H^* = H(s, { \bf{{{z}}^{ \star}}}( s , {\bf{{z}}_0}), \mu)- H(s, {\bf{{\tilde{z}}^{ \star}}}( s , {\bf{\tilde{z}}_0}), \mu )$.

Now,
\begin{eqnarray}
\norm{\Delta {\bf z}}  & \leq &   Ke^{- \gamma t} \norm{ {\bf{{z}}_0}-{\bf{\tilde{z}}_0} } + \nonumber \\ 
& & \int_{0}^{t} KL  e^{- \gamma(t-s)} \norm{  { \bf{{{z}}^{ \star}}}( s , {\bf{{z}}_0})-  {\bf{{\tilde{z}}^{ \star}}}( s , {\bf{\tilde{z}}_0})} ds.
\end{eqnarray}

By multiplying both members of the previous inequality by $e^{\gamma t}$, we obtain

\begin{eqnarray}
e^{\gamma t}\norm{\Delta {\bf z}} & \leq &  K \norm{ {\bf{{z}}_0}-{\bf{\tilde{z}}_0} } + \nonumber \\ 
& & \int_{0}^{t} KL  e^{ \gamma s} \norm{  { \bf{{{z}}^{ \star}}}( s , {\bf{{z}}_0})-  {\bf{{\tilde{z}}^{ \star}}}( s , {\bf{\tilde{z}}_0}) } ds.
\end{eqnarray}

For this inequality we have, with $M= K \Vert {\bf{{z}}_0}-{\bf{\tilde{z}}_0} \Vert$, $f(t)= e^{\gamma t}\Vert{ \bf{{{z}}^{ \star}}}( t , {\bf{{z}}_0} ) - {\bf{{\tilde{z}}^{ \star}}}( t , {\bf{\tilde{z}}_0} )\Vert$ and $g(t)= KL$, the hypotheses of Gronwall's lemma. Therefore, we can conclude that
\begin{equation*}
e^{\gamma t}\norm{\Delta {\bf z}} \leqslant K \norm{ {\bf{{z}}_0}-{\bf{\tilde{z}}_0} } e^{\int_{0}^{t} KL ds}, \quad t \geqslant 0.
\end{equation*}

Thus,
\begin{equation} \label{interesting}
\norm{\Delta {\bf z}} \leqslant K  e^{\left(KL - \gamma \right)t } \norm{ {\bf{{z}}_0}-{\bf{\tilde{z}}_0} },\quad t \geqslant 0,
\end{equation}
because in the \textit{Theorem 1}, $\displaystyle\frac{KL}{\gamma} \leqslant \frac{1}{4}$. Relation $(\ref{interesting})$ implies the result of uniform continuity with respect to $t$. Moreover, $\norm{{ \bf{{{z}}^{ \star}}}( t , {\bf{{z}}_0} ) - {\bf{{\tilde{z}}^{ \star}}}( t , {\bf{\tilde{z}}_0} )}$ approaches zero exponentially as $t \rightarrow \infty$ and ${\norm{{ \bf{{{z}}^{ \star}}}( \cdot , {\bf{{z}}_0} ) - {\bf{{\tilde{z}}^{ \star}}}( \cdot , {\bf{\tilde{z}}_0} )}}_{\infty} \leqslant  K  \norm{ {\bf{{z}}_0}-{\bf{\tilde{z}}_0} }  $. \\

\section{Numerical results}
We have selected as an example, an archetype of a piece-wise linear chaotic systems. This class of systems, in addition to simplifying the analytical treatment of problems, has a large number of practical applications. Among them, those associated with cryptography based on syncronization piecewise-linear chaotic systems such as \cite{Guillen} stand out, where an attempt is made to improve the security of digital communications, so necessary in these times.

Chua's circuit is a bridge to connect the characteristic associated to dynamics of nonlinear phenomena as: stable orbits, bifurcations or attractors, with the study of  experimental chaos. This circuit, in its classic configuration, is one of the simplest chaotic systems containing an inductor, two capacitors which are the linear energy-storage elements, a linear resistor and one 2-terminal nonlinear resistor characterized by a current-voltage  characteristic which has a negative slope \cite{Irimiciuc2015}. 
All circuit elements are passive except for the nonlinear resistor; this element must be active in order for the circuit to become chaotic \cite{Chua1992}. 

In this work, we will use the adimensional form of equation system which is a set of interdependent equations in the form of a 3-dimensional autonomous piece-wise linear ordinary differential  equation (flow) described by 
\begin{equation}\label{Cheq}
\begin{cases}
\Dot{x}&=\alpha(y-x-f(x)),\\
\Dot{y}&= x-y+z,\\
\Dot{z}&=-\beta y, 
\end{cases}
\end{equation}
where 
\begin{equation*}
f(x)= bx+\frac{1}{2}(b-a)\left[\lvert x+1\rvert-\lvert x-1\rvert\right],
\end{equation*}
and $a, b, \alpha$ and $\beta$ are real parameter, $a<b<0,~ \alpha>0$ and $\beta >0$. The system $(\ref{Cheq})$ is known as Chua's equation.

This piecewise function can be written as follows
\begin{equation}\label{f(x)}
f(x)=
\begin{cases}
bx-(a-b), &\text{if } x \leq -1,\\
ax, &\text{if }  -1\leq x\leq 1,\\
bx+(a-b), &\text{if } x\geq 1.
\end{cases}
\end{equation}

This system is the honorary member of a class of chaotic systems to which also belong: Sprott systems \cite{sprott}, Murali-Lakshmanan-Chua circuits \cite{LAKSHMANAN}, memristor based circuits as in \cite{Deng} or 3-dimensional piecewise-linear system as shown in \cite{Yang,Elhadj}.  That systems can be relatively easily designed through electronic circuits and the corresponding
dynamical equations can be numerically analysed and theoretically investigated. 

In order to apply the theoretical results in the case of Chua's sytems $(\ref{Cheq})$, we consider

\begin{equation*}
\mathbf{z}=
\begin{pmatrix}
\bar{x} \\
\bar{y} \\
\bar{z}
\end{pmatrix},
\quad {\bf x}(t, {\bf x}_0, \bar{\mu})=
\begin{pmatrix}
x \\
y\\
z
\end{pmatrix},
\end{equation*}

\noindent
and the parameters,

\begin{equation*}
\quad \mu = 
\begin{pmatrix}
\alpha\\
\beta
\end{pmatrix}~~ and ~~
\bar{\mu} =
\begin{pmatrix}
\bar{\alpha}\\
\bar{\beta}
\end{pmatrix}.
\end{equation*}

We have to find the representation of $(\ref{zpuntito})$ for the case of $(\ref{Cheq})$. For that, first let us compute ${\bf f}( \mathbf{z}+ {\bf x}(t, {\bf x}_0,\bar{ \mu}) , \mu)- {\bf f}\left( {\bf x}(t, {\bf x}_0, \bar{\mu} ) , \bar{\mu}\right)$. Then,  
\begin{equation*}
{\bf f}( \mathbf{z}+ {\bf x}(t, {\bf x}_0, \bar{\mu}), \mu)=  \begin{pmatrix}
\alpha((\bar{y}+y)-(\bar{x}+x)-f(\bar{x}+x))\\
(\bar{x}+x)-(\bar{y}+y)+(\bar{z}+z)\\
-\beta (\bar{y}+y)
\end{pmatrix},
\end{equation*}
so that, ${\bf f}( \mathbf{z}+ {\bf x}(t, {\bf x}_0,\bar{ \mu}) , \mu)- {\bf f}\left( {\bf x}(t, {\bf x}_0, \bar{\mu} ) , \bar{\mu}\right)$ results in
\begin{align*}
\begin{pmatrix}
\alpha(\bar{y}-\bar{x})\\
\bar{x}-\bar{y}+\bar{z}\\
-\beta \bar{y}
\end{pmatrix}+\begin{pmatrix}
(\alpha-\bar{\alpha})(y-x-f(x))\\
0\\
(\bar{\beta}-\beta)y
\end{pmatrix}
\\+
\begin{pmatrix}
\alpha(f(x)-f(\bar{x}+x))\\
0\\
0
\end{pmatrix}.
\end{align*} 

Rewriting the last expression we have
\begin{align*}
& \begin{pmatrix}
-\alpha & \alpha &0\\
1 &-1 & 1\\
0 &-\beta & 0
\end{pmatrix}
\begin{pmatrix}
\bar{x}\\\bar{y}\\ \bar{z} 
\end{pmatrix}
+
\begin{pmatrix}
(\alpha-\bar{\alpha})(y-x-f(x))\\
0\\
(\bar{\beta}-\beta)y
\end{pmatrix}
\\&+
\begin{pmatrix}
\alpha(f(x)-f(\bar{x}+x))\\
0\\
0
\end{pmatrix}.
\end{align*}

Now, setting  
\begin{equation}\label{MatrixAc}
A=A(\mu):=
\begin{pmatrix}
-\alpha & \alpha &0\\
1& -1 &1\\
0 &-\beta &0
\end{pmatrix},
\end{equation}
\begin{equation}\label{FGc}
G(t, \mu, \bar{\mu}):=
\begin{pmatrix}
(\alpha-\bar{\alpha})(y -x-f(x))\\
0\\
(\bar{\beta}-\beta)y
\end{pmatrix},
\end{equation}
and
\begin{equation}\label{MatrixHc}
H(t, \mu, \mathbf{z})=
\begin{pmatrix}
\alpha(f(x)-f(\bar{x}+x))\\
0\\
0
\end{pmatrix},
\end{equation}
\\
the system (\ref{2master})-(\ref{2slave}), that through the transformation (\ref{Transformation}) lead us to equation (\ref{NonAut}), in the context of Chua's equations $(\ref{Cheq})$, becomes
\begin{align}
\mathbf{\dot{z}}&=  \nu \mathbf{z} + A\mathbf{z} + G(t, \mu, \bar{\mu})+H(t, \mu, \mathbf{z})\notag\\
&= (A+\nu I)\mathbf{z}+ G(t, \mu, \bar{\mu})+H(t, \mu, \mathbf{z}).\label{zpcq}
\end{align} 

Notice that we have obtained an expression like the one given in $(\ref{Decomposition})$. Our next goal is to establish that for the expression given in  $(\ref{zpcq})$ the hypothesis $H_1$, $H_2$ and $H_3$ are satisfied. To prove $H_1$, we consider $B=A+\nu I$ where $B \in \mathcal{M}_{3\times3}$ so that, it is important to know some facts about the matrix $A$. For any $\alpha$, $\beta \in \mathbb{R}$, the eigenvalues of $A$ are given by 

\begin{align*}
\lambda_{1_A}=&\left( -\frac{1}{2}-\displaystyle\frac{\sqrt{3} i}{2}\right) \, \mathit{h_2(\alpha, \beta)}\\&-\frac{\left( \frac{\sqrt{3} i}{2}-\displaystyle\frac{1}{2}\right) \, \left( \frac{\beta }{3} -\frac{ {{\left( \alpha +1\right) }^{2}}}{9}\right) }{\mathit{h_2(\alpha, \beta)}}+\frac{\left(-1\right) \, \left( \alpha +1\right) }{3},
\\\lambda_{2A}=&\left(-\frac{1}{2} +\frac{\sqrt{3} i}{2}\right) \, \mathit{h_2(\alpha, \beta)}\\&-\frac{\left( -\frac{1}{2}-\frac{\sqrt{3} i}{2}\right) \, \left( \frac{\beta }{3}+\frac{\left( -1\right) \, {{\left( \alpha +1\right) }^{2}}}{9}\right) }{h_2(\alpha, \beta)}-\frac{ \left( \alpha +1\right) }{3},
\\\lambda_{3A}=&\mathit{h_2(\alpha, \beta)}-\displaystyle\frac{\frac{\beta }{3}+\frac{\left( -1\right) \, {{\left( \alpha +1\right) }^{2}}}{9}}{\mathit{h_2(\alpha, \beta)}}-\frac{ \left( \alpha +1\right) }{3},
\end{align*}
where
\begin{equation*}
h_2(\alpha, \beta)={{\left( \mathit{h_1(\alpha, \beta)}+\frac{\left( \alpha +1\right)  \beta -3 \alpha  \beta }{6}-\frac{ {{\left( \alpha +1\right) }^{3}}}{27}\right) }^{\frac{1}{3}}},
\end{equation*}
and

	\begin{equation*}
	h_1(\alpha, \beta)= \frac{\sqrt{\beta \, \left( 4 {{\beta }^{2}}+\left( 8 {{\alpha }^{2}}-20 \alpha -1\right)  \beta +4 {{\alpha }^{4}}+12 {{\alpha }^{3}}+12 {{\alpha }^{2}}+4 \alpha \right) }}{2 {{3}^{\frac{3}{2}}}}.
	\end{equation*}

\begin{proposition}
	$\lambda$ is an eigenvalue of  $A$ if and only if $\lambda + \nu$ is an eigenvalue of $B$. 
\end{proposition}

From here, we consider $\alpha=9$ and $\beta=\frac{100}{7}$ so, the eigenvalues of $A$ are 
\begin{align*}
\lambda_{1A}&=-0.0639-3.608i,\\
\lambda_{2A}&=-0.0639+3.608i,\\
\lambda_{3A}&=-9.8721.
\end{align*}
Now, the proposition 3 implies that if $\nu < 0.064$, then,  for $\alpha=9$ and $\beta=\frac{100}{7}$, all the eigenvalues of the matrix $B$ have negative real part. Thus, $H_1$ is satisfied.

Next, we consider  $G(t, \mu, \bar{\mu})$ given by $(\ref{FGc})$. Recall that 
\begin{equation*}
{\bf x}(t, {\bf x}_0, \bar{\mu})=
\begin{pmatrix}
x \\
y\\
z
\end{pmatrix}
\end{equation*}
is bounded. This implies that $x,y $, $f(x)$ are bounded and there exists $R>0$ such that $| y-x +f(x) | \leq R$ and $|y|\leq R$ for all $t \geq 0$. The next proposition tell us that $H_2$ is satisfied.
\begin{proposition}
	Given $\varepsilon>0$, then 
	\begin{equation*}
	\norm{ G(t, \mu, \Bar{\mu}}<\varepsilon
	\end{equation*}
	for $t\geq 0$ if $\norm{\mu-\Bar{\mu}}<\frac{\epsilon}{R}$ holds.
\end{proposition}

Now, hypothesis $H_3$ is satisfied due to the fact that
\begin{proposition}
	The function $H(t, \mu, \bold{z})$ given by $(\ref{MatrixHc})$ is globally Lipschitz in $\mathbf{z}$. Moreover, the Lipschitz constant $L$ can be chosen as $L= \alpha | a |$, being $a$ as in $(\ref{f(x)})$.
\end{proposition}

We say that the Chua's System fulfills the conditions $H_1$, $H_2$ and $H_3$. Summarizing for $\alpha= 9, \beta=\frac{100}{7}, a=-\frac{8}{7}$ and $b=-\frac{5}{7}$, we found constants $K, \gamma$, and $ L$ such that 
\begin{equation*}
K=5.8472, \quad \gamma=9.936 - 2\nu\quad\text{and}\quad L=9\mid a\mid=\frac{72}{7}.
\end{equation*} 

Thus, for values of $\nu_c < -1.79$, the condition $\frac{KL}{\gamma}<\frac{1}{4}$ is satisfied. Therefore, we can apply the \textit{Theorem 1} so that
\begin{equation*}
\norm{\mathbf{y}- \mathbf{x}(t, x_0, \bar{\mu})}\leq \rho.
\end{equation*}

The technique exposed in the system  $(\ref{2master})-(\ref{2slave})$ considers two similar copies of the system to be synchronized with different initial conditions on them. Let $\mathbf{x}=(x, y, z)\in \mathbb{R}^3$ be the generalized coordinates corresponding to the master system, and $\mathbf{y}=(x_s, y_s, z_s)\in \mathbb{R}^3$ those of the slave system. Thus, the whole system is
\begin{equation*}
\begin{cases}
\dot{x}= \Bar{\alpha}(y-x-f(x)),
\\\dot{y}=x-y+z,
\\\dot{z}= -\Bar{\beta} y,
\\\dot{x}_s=\alpha(y_s-x_s-f(x_s))+\nu(x_s-x), 
\\\dot{y}_s=x_s-y_s+z_s+\nu(y_s-y),
\\\dot{z}_s=-\beta y_s+\nu(z_s-z).
\end{cases}
\end{equation*}

As we mentioned before, we concentrate our attention on the master$-$slave system with the usual parameters $\Bar{\mu}=(\Bar{\alpha}, \Bar{\beta})=\left( 9, \frac{100}{7}\right)$, and $(a, b)=\left(-\frac{8}{7}, -\frac{5}{7}\right)$. Let us choose $\nu = -2$ since it corresponds to the minor number respect to the negative real part of the eigenvalues mentioned in \textit{Proposition 7}. Also, let us consider a vector $\mu=(\alpha, \beta)=\left(8.9, \frac{99}{7}\right)$, closer to $\bar{\mu}$.

\begin{figure}[ht]
	\begin{center}
    \includegraphics[width=8cm]{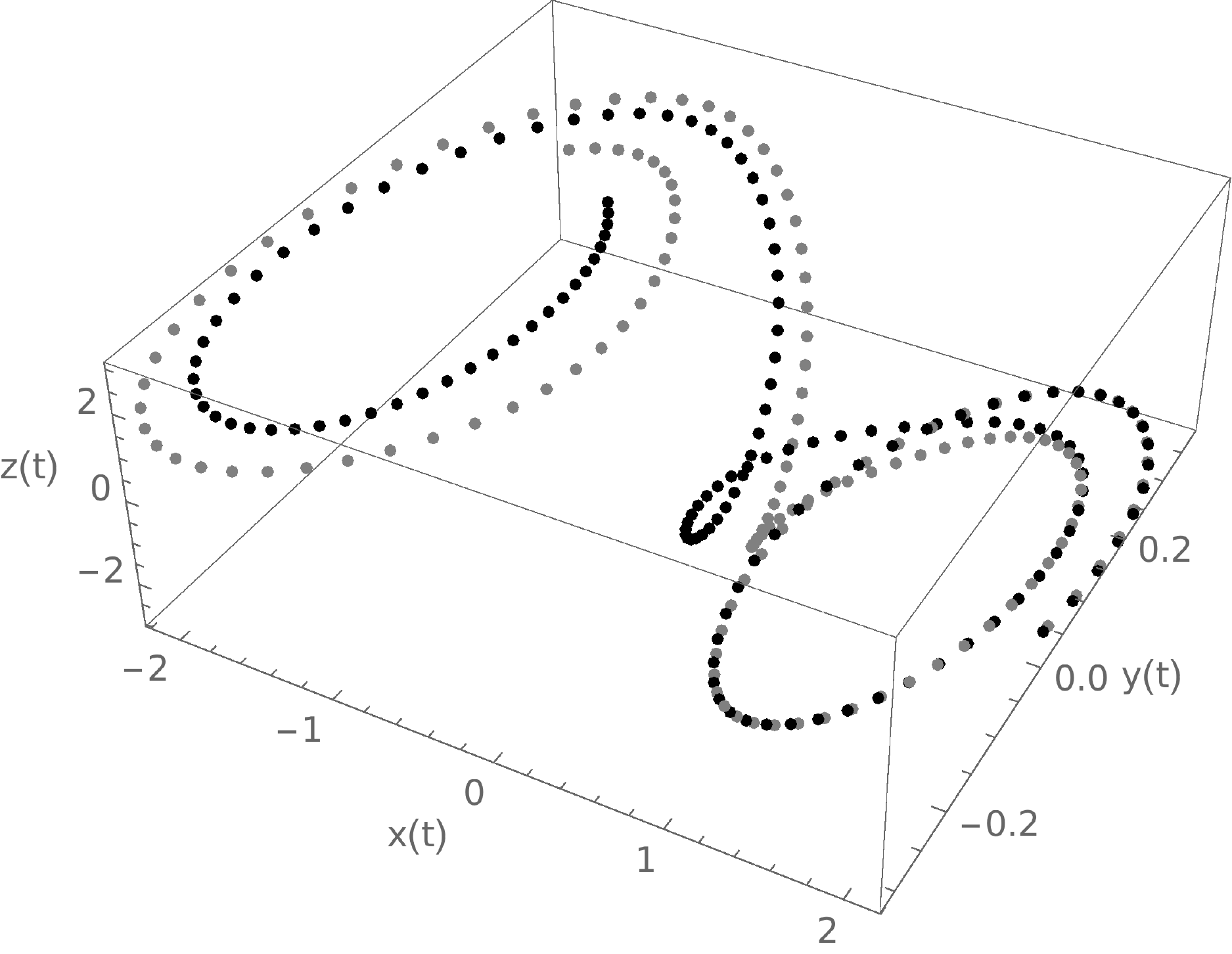}
	\caption{Evolution of the master and slave systems from diferents initial conditions. The gray points represent the orbit of the master and the black ones that of the slave system.}
	\label{catractor}
    \end{center}
\end{figure}

To show the performance of the synchronization technique, we starting from two slightly initial conditions $(0.2, 0.4, 1.0)$ and $(0.2, 0.5, 0.9)$ and show in Figure \ref{catractor} the evolution of the systems.

A quality criteria for synchronization can be settled down by defining  the {\it synchronization error} ($E$) in the interval $(t_0,T)$, as.

\begin{equation}
E(\nu) = \int_{t_0}^{T} \parallel ({\bf x}(t)-{\bf y}(t) \parallel^2 dt
\end{equation}

\noindent
The following figure shows the dependence of the error of synchronization as a function of the  coupling parameter $\nu$.

	\begin{figure}[ht]
		\begin{center}
		\includegraphics[width=8cm]{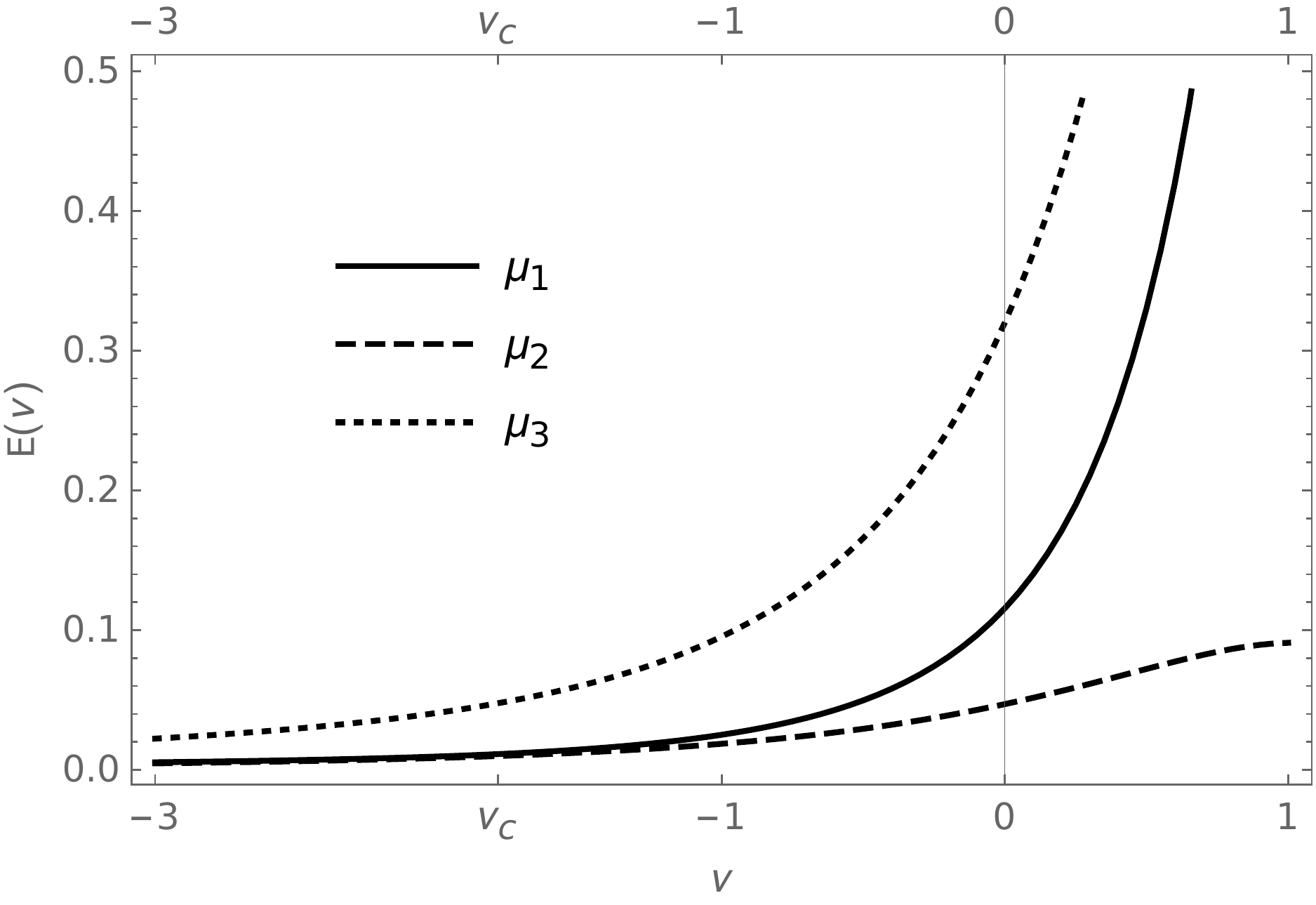}
		\caption{Error of synchronization as a function of the coupling parameter, in the case of several sets of values of the  slave system $\mu_1 = \left( 9, \frac{95}{7}\right)$, $\mu_2 = \left(8.5, \frac{100}{7}\right)$, $\mu_3 = \left( 8, \frac{95}{7}\right)$.}
		\label{error}
	   \end{center}
	\end{figure}

\section{Concluding remarks}

In this work, a synchronization strategy has been proposed and developed for a class of piecewise-linear chaotic systems.

Our findings can be summarized in the following list:

\begin{itemize}
	\item Our approach appears to be useful in inducing synchronization in 
	the case of a class of systems with many practical applications, among which cryptography or the training of reaction-diffusion neural networks stand out.
	\item The scheme allows an analytical development of the strategy, which in turn gives us ideas for solving the problem in other contexts.
	\item The main result  to calibrate the intensity of the coupling, given the difference between the parameters.
\end{itemize}

The presentation exposes explicitly and in considerable detail not only the results, but the proofs in a way that we hope will also be pedagogical. 

Finally, although the study of the problem is outside the scope of the work, we believe that it is possible to extend this result to the case of continuous chaotic systems (ODEs) whose non-linearity is approximated by piecewise linear functions.

\newpage

\newpage

\section*{Appendix}
\appendix
In the following, the proofs of the propositions, lemmas and theorems previously presented are shown, in the cases in which it is necessary.

\section{Proof of Proposition 1}
\begin{proof}
	The proof of this result is strongly based on the Jordan canonical form of the matrix $B$ and can be seen in \cite{Hale1980} \textit{(Theorem 4.2., (ii))}.
\end{proof}

\section{Proof of Lemma 1}
\begin{proof}
	It is a direct consequence of the main Theorem of Calculus.
\end{proof}

\section{Proof of Proposition 2}
\begin{proof}
	We are going to show that $X-\mathcal{G}\left(\mathbf{z}_{0}, \rho, \mu, \bar{\mu}\right)$ is open i.e. given $ \bar{\bf{z}} \in X-\mathcal{G}\left(\mathbf{z}_{0}, \rho, \mu, \bar{\mu}\right) $ there exists an $\varepsilon-$neighborhood of $\bar{\bf{z}}$ which is contained in  $X-\mathcal{G}\left(\mathbf{z}_{0}, \rho, \mu, \bar{\mu}\right)$. If $ \bar{\bf{z}} \in X-\mathcal{G}\left(\mathbf{z}_{0}, \rho, \mu, \bar{\mu}\right) $ then $ \bar{\bf{z}}(0) \neq \mathbf{z}_{0}$ or for some $\bar{t} \geq 0$ we have that $\norm{\bar{\bf{z}}(\bar{t})}>\rho$. 
	
	If $\bar{\bf{z}}(0) \neq \bf{z}_{0}$, then $\norm{\bar{\bf{z}}(0)-\bf{z}_{0}}:=\bar{r}>0 .$ Choose
	$\varepsilon=\frac{\bar{r}}{2}$ and consider the set
	\begin{equation*}
	\left\{ {\bf{z}} \in C_{b}\left([0, \infty), \mathbb{R}^{n}\right):\| \textbf{z}-\bar{\mathbf{z}}\|_{\infty}<\varepsilon\right\}.
	\end{equation*}
	
	We have that this $\varepsilon$ -neighborhood is contained in  $X-\mathcal{G}\left(\mathbf{z}_{0}, \rho, \mu, \bar{\mu}\right)$. In fact: If $ {\bf{z}} \in \left\{ {\bf{z}} \in C_{b}\left([0, \infty), \mathbb{R}^{n}\right):\| \textbf{z}-\bar{\mathbf{z}}\|_{\infty}<\varepsilon\right\} $, then 
	\begin{align*}
	\norm{\mathbf{z}(0)-\mathbf{z}_{0}} &=\norm{\mathbf{z}(0)-\bar{\mathbf{z}}(0)+\bar{\mathbf{z}}(0)-\mathbf{z}_{0}} \\
	& \geqslant\norm{\bar{\mathbf{z}}(0)-\mathbf{z}_{0}}-\|\mathbf{z}(0)-\bar{\mathbf{z}}(0)\| \\
	&=\bar{r}-\frac{\bar{r}}{2}>0 .
	\end{align*}
	
	Thus, ${\bf{z}}(0) \neq \bf{z}_0$. 
	
	If $\|\bar{\mathbf{z}}(\bar{t})\|> \rho$ for some $ \bar{t} \geqslant 0$, then the
	$\varepsilon$ -neighburhood of $\bar{\mathbf{z}}$, with $\varepsilon=\| \bar{\mathbf{z}}(\bar{t})\|-\rho$, is contained in  $X-\mathcal{G}\left(\mathbf{z}_{0}, \rho, \mu, \bar{\mu}\right)$. In fact: If $ {\bf{z}} \in \left\{ {\bf{z}} \in C_{b}\left([0, \infty), \mathbb{R}^{n}\right):\| \textbf{z}-\bar{\mathbf{z}}\|_{\infty}<\varepsilon\right\} $, then
	\begin{align*}
	\norm{\mathbf{z}(\bar{t})} &=\norm{ \mathbf{z}(\bar{t})-\bar{\mathbf{z}}(\bar{t})+\bar{\mathbf{z}}(\bar{t}) }\\
	& \geqslant\norm{\bar{\mathbf{z}}(\bar{t})}-\norm{\mathbf{z}(\bar{t}) -\bar{\mathbf{z}}(\bar{t}) }\\
	&=\varepsilon+\rho- \norm{\mathbf{z}(\bar{t}) -\bar{\mathbf{z}}(\bar{t}) }  \\
	&>\rho .
	\end{align*}
	
	Thus, $ \norm{ {\bf{z}}(\bar{t})} > \rho$.
\end{proof}

\section{Proof of Lemma 2}

\begin{proof}
	Define $ h(t)=M+\int_{a}^{t} f(s)g(s) d s$. We have that $h(a)=M$ and $\dot{h}(t)=f(t) g(t)$. Now, since $f(t)\leqslant h(t) $, $ f(t) \geqslant 0$ and $ g(t) \geqslant 0$ for $a \leqslant t \leqslant b$ it is obtained 
	\begin{equation*}
	\dot{h}(t)=f(t) g(t)\leqslant h(t)g(t)~. 
	\end{equation*}
	
	By multiplying both members of this inequality by $e^{- \int_{a}^{t}g(s)ds}$, we obtain 
	\begin{equation*}
	e^{- \int_{a}^{t}g(s)ds}\dot{h}(t) \leqslant e^{- \int_{a}^{t}g(s)ds} h(t)g(t)~.
	\end{equation*}
	
	Thus, 
	\begin{equation*}
	e^{- \int_{a}^{t}g(s)ds}\left( \dot{h}(t) - h(t)g(t) \right) \leqslant 0 ~~ \mbox{and} ~~ \frac{d}{dt}\left( e^{- \int_{a}^{t}g(s)ds} h(t) \right) \leqslant 0~.
	\end{equation*}
	
	Now, integrating from $a$ to $t$ we get
	\begin{equation*}
	e^{- \int_{a}^{t}g(s)ds}h(t) - e^{- \int_{a}^{t}g(s)ds} h(a) \leqslant 0~,
	\end{equation*}
	which implies $e^{- \int_{a}^{t}g(s)ds}h(t) \leqslant h(a)=M$ and $ h(t) \leqslant Me^{\int_{a}^{t}g(s)ds} $. Finally, the fact that $f(t) \leqslant h(t)$ produces the result
	\begin{equation*}
	f(t) \leqslant M e^{ \int_{a}^{t}g(s)ds}.
	\end{equation*}
\end{proof}

\section{Proof of Theorem 1}

\begin{proof}
	Let  ${\bf{z}} \in \cal G \left(  \bf{z}_{0} , \rho , \mu , \bar{\mu}  \right)$. From $H_1$,  (\ref{Condition2}), (\ref{otraC}), we obtain
	\begin{align*}
	\norm{ \left( T {\bf{z}}  \right)(t)}
	~\leq & \norm{e^{Bt}{\bf{z}}_0}  + \int_{0}^{t} \norm{ e^{B(t-s)}N(s, {\bf{z}}(s), \mu , \bar \mu)} ds    
	\\ \leq &  Ke^{- \gamma t} \norm{\bf{z}_0}+K \int_{0}^{t}e^{- \gamma (t-s)} \norm{N(s , {\bf{z}}(s), \mu , \bar \mu)} ds 
	\\\leq & Ke^{- \gamma t} \norm{{\mathbf{z}}_0}  +K \int_{0}^{t}e^{- \gamma (t-s)} \norm{\left(  G(s , \mu, \bar{\mu} ) \right)} ds 
	\\& +K \int_{0}^{t}e^{- \gamma (t-s)} \norm{\left(   H(s , {\mathbf{z}}(s), \mu ) \right)} ds
	\\\leq & \frac{\rho}{2} + \frac{\rho \gamma}{4} \int_{0}^{t}e^{- \gamma (t-s)} ds +KL \int_{0}^{t}e^{- \gamma (t-s)} \norm{  {\mathbf{z}}(s)} ds 
	\\\leq &  \frac{\rho}{2} + \frac{\rho \gamma}{4} \int_{0}^{t}e^{- \gamma (t-s)} ds +KL \norm{{\bf{z}} }_{\infty} \int_{0}^{t} e^{- \gamma (t-s)} ds  
	\\ = &\frac{\rho}{2} +  \left( \frac{\rho \gamma}{4} + KL \norm{  {\bf{z}}}_{\infty} \right) \frac{1}{\gamma}(1- e^{- \gamma t}) 
	\\ \leq& \frac{\rho}{2} +  \left( \frac{\rho \gamma}{4} + KL \norm{  {\bf{z}}}_{\infty}\right) \frac{1}{\gamma}  \\ = & \frac{\rho}{2} + \frac{\rho}{4}+ \frac{KL}{\gamma}  \norm{  {\bf{z}}}_{\infty}
	\\ \leq & \rho.
	\end{align*} 
	
	Thus, 
	$\norm{T{\bf{z}}} \leq \rho $ and $ T{\bf{z}} \in \cal G \left(  {\bf{z}}_{0} , \rho , \mu , \bar{\mu}  \right)$. 
	
	Now, let us take  ${\bf{z_1}},{\bf{z_2}} \in \cal G \left(  {\bf{z}}_{0} , \rho , \mu , \bar{\mu}  \right)$, the same type of estimates yields
	\begin{align*}
	\norm{\left( T {\bf{z_2}}  \right)(t) - \left( T {\bf{z_1}}  \right)(t)  }& \leq \int_{0}^{t} \norm{ e^{B(t-s)} \Delta H }ds     
	\\  & \leq \int_{0}^{t}Ke^{- \gamma (t-s)} \norm{\Delta H }ds 
	\\&  \leq  \int_{0}^{t}KLe^{- \gamma (t-s)} \norm{ {\bf{z_2}}(s)- {\bf{z_1}}(s)} ds  
	\\& \leq \left( \int_{0}^{t}KLe^{- \gamma (t-s)}ds  \right){ \norm{{\bf{z_2}}- {\bf{z_1}}}_{\infty}} 
	\\& =  \frac{KL}{\gamma} \left( 1 - e^{- \gamma t} \right) { \norm{ {\bf{z_2}}- {\bf{z_1}}}_{\infty}}  
	\\ & \leq \frac{KL}{\gamma}{ \norm{{\bf{z_2}}- {\bf{z_1}}}_{\infty}} 
	\\& \leq \frac{1}{4} { \norm{{\bf{z_2}}- {\bf{z_1}}}_{\infty}}.
	\end{align*} 
	
	\noindent
	with $\delta H = H(s , {\bf{z_2}}(s), \mu )- H(s , {\bf{z_1}}(s), \mu )$
	
	Thus, $T$ is a contraction on  $\cal G \left( {\bf{z}}_{0} , \rho , \mu , \bar{\mu}  \right)$ and there is a unique fixed in  $\cal G \left( {\bf{z}}_{0} , \rho , \mu , \bar{\mu}  \right)$.
\end{proof}

\section{Proof of Proposition 3}
\begin{proof}
	
	Let $B$ be a $3\times 3$ real matrix given by 
	\begin{equation}
	B= A+\nu I=\begin{pmatrix}
	-\alpha+\nu &\alpha&0\\
	1&-1+\nu&1\\
	0& -\beta&\nu
	\end{pmatrix}.
	\end{equation}
	
	By \textit{Proposition 1}, for
	any value of $\alpha, \beta $, the eigenvalues of B are given by
	\begin{equation*}
	\lambda_{1B}= \lambda_{1A}+\nu, \quad \textit{i}=1,2,3,
	\end{equation*}
	then for values of $\alpha=9$ and $\beta=\frac{100}{7}$, the eigenvalues become 
	\begin{align*}
	\lambda_{1B}&=-0.0639+\nu-3.6082i,\\
	\lambda_{2B}&=-0.0639+\nu+3.6082i,\\
	\lambda_{3B}&=-9.8721+\nu.
	\end{align*}
	
	By choosen $\nu< 0.064$, we have estabished that all the eigenvalues have negative real part. By  \textit{Proposition 3}, $B$ has the descomposition  
	\begin{equation*}
	B=PJP^{-1},
	\end{equation*}
	where $J$ is given by
	\begin{equation*}
	J=\begin{pmatrix}
	-0.0639+\nu & -3.6082&0\\
	3.6082 & -0.0639+\nu &0\\
	0&0&-9.8721+\nu
	\end{pmatrix}.
	\end{equation*}
	
	By \textit{Proposition 2}, the eigenvalues of $A$ are eigenvalues for $B$ so $P$ becomes
	\begin{equation*}
	P=\begin{pmatrix}
	1&0&1\\
	0.9929&-0.4009& -0.0969\\
	-1.5172&-3.9579&-0.1402
	\end{pmatrix},
	\end{equation*}
	the inverse corresponding to this matrix is 
	\begin{equation*}
	P^{-1}=\begin{pmatrix}
	0.06728&	0.8135&	-0.0824\\
	-0.0588&	-0.2830&	-0.2240\\
	0.9327&	-0.8135&	0.0824
	\end{pmatrix}.
	\end{equation*}
	
	We know that 
	\begin{equation}
	e^{Bt}=P e^{Jt}P^{-1}.
	\end{equation}
	
	By taking the Euclidean norm in both sides, we get 

		\begin{align*}
		\norm{e^{Bt}}&=\norm{Pe^{Jt}P^{-1}}\\
		&\leq \norm{P}\norm{P^{-1}}\norm{e^{Jt}}\\ 
		&=\norm{P}\norm{P^{-1}}\norm{\begin{pmatrix}
			e^{(-0.0639+\nu)t}\cos{(3.6082t)}&-e^{-0.0639t}\sin{(3.6082t)}&0\\
			e^{-0.0639t}\sin{(3.6082t)}&e^{(-0.0639+\nu)t}\cos{(3.6082t)}&0\\
			0&0&e^{(-9.8721+\nu)t}
			\end{pmatrix}}\\
		&\leq 5.8470 \left(e^{(-9.936 + 2\nu)t}\right)
		\\&=Ke^{-\gamma t},  
		\end{align*}

	\noindent
	where $K= 5.8470$ and  $\gamma=9.936 - 2\nu$.
\end{proof}

\section{Proof of Proposition 4}

\begin{proof}
	Let $\varepsilon>0$ be arbitrary. We have
	$$
	\begin{aligned}
	\|G(t, \mu, \bar{\mu})\|^{2} &=|\alpha-\bar{\alpha}|^{2}|y-x-f(x)|^{2} \\
	&+|\beta-\bar{\beta}|^{2}|y|^{2} \\
	& \leq R^{2}\|\mu-\bar{\mu}\|^{2}.
	\end{aligned}
	$$
	Therefore $\|G(t, \mu, \bar{\mu})\| \leqslant R\|\mu-\bar{\mu}\|$ and if $\|\mu-\bar{\mu}\|<\frac{\varepsilon}{R}$ then the desired result is obtained.
\end{proof}

\section{Proof of Proposition 5}

\begin{proof}
	Consider 
	\begin{equation*}
	\mathbf{z_1}=
	\begin{pmatrix}
	\bar{x}_1 \\
	\bar{y}_1 \\
	\bar{z}_1
	\end{pmatrix},
	\mathbf{z_2}=
	\begin{pmatrix}
	\bar{x}_2 \\
	\bar{y}_2\\
	\bar{z}_2
	\end{pmatrix} ~ \mbox{be in} ~ \mathbb{R}^{3}.
	\end{equation*}
	
	\noindent
	We have
	$$H(t, \mu, \mathbf{z_2})-H(t, \mu, \mathbf{z_1})= \begin{pmatrix}
	\alpha(f(\bar{x}_1+x) - f(\bar{x}_2+x))\\0\\0
	\end{pmatrix}$$
	
	\noindent
	From the definition of $f$, given in (\ref{f(x)}), we pay attention, according to the values of  $\bar{x}_1+x$ and $\bar{x}_2+x$, to nine cases:
	\begin{enumerate}[i)]
		\item $ \bar{x}_1 + x \leq -1$, ~$ \bar{x}_2 + x \leq -1$  
		\item $\bar{x}_1+x \leq -1$, ~ $\mid\bar{x}_2+x\mid \leq 1$
		\item $ \bar{x}_1 + x \leq -1$, ~$ 1 \leq \bar{x}_2 + x $ 
		\item $\mid\bar{x}_1+x\mid \leq 1$, ~ $ \bar{x}_2+x \leq -1$
		\item  $\mid\bar{x}_1+x\mid \leq 1$, ~ $\mid\bar{x}_2+x\mid \leq 1  $
		\item  $\mid\bar{x}_1+x\mid \leq 1$, ~ $ 1 \leq \bar{x}_2+x $
		\item  $ 1 \leq \bar{x}_1 + x $, ~$ \bar{x}_2 + x \leq -1$ 
		\item  $ 1 \leq \bar{x}_1+x $, ~ $\mid\bar{x}_2+x\mid \leq 1$
		\item   $ 1 \leq \bar{x}_1 + x $, ~$ 1 \leq \bar{x}_2 + x $  .
	\end{enumerate}
	Cases i) and ix) produce
	$$\alpha(f(\bar{x}_1+x) - f(\bar{x}_2+x))= \alpha b (\bar{x}_1 - \bar{x}_2).$$
	For case v) it is obtained
	$$\alpha(f(\bar{x}_1+x) - f(\bar{x}_2+x))= \alpha a (\bar{x}_1 - \bar{x}_2).$$
	For case ii) we have that $\bar{x}_2-\bar{x}_1 \geq 0$ and 
	$$f(\bar{x}_1+x) - f(\bar{x}_2+x)= (b-a)(x+1)+b \bar{x}_1 -a \bar{x}_2 .$$
	Now, conditions $a<b<0$ and $x+1 \leq - \bar{x}_1 $ imply that 
	\begin{equation}\label{h1}
	f(\bar{x}_1+x) - f(\bar{x}_2+x) \leq  a (\bar{x}_1 - \bar{x}_2 ).
	\end{equation}
	Similarly, from conditions $a<b<0$ and $x+1 \geq - \bar{x}_2$ it is obtained
	\begin{equation}\label{h2}
	f(\bar{x}_1+x) - f(\bar{x}_2+x) \geq  b (\bar{x}_1 - \bar{x}_2 ).
	\end{equation}
	From (\ref{h1}) and (\ref{h2}), we may conclude that
	$$ | \alpha (	f(\bar{x}_1+x) - f(\bar{x}_2+x)) | \leq \alpha |a| |\bar{x}_2 - \bar{x}_1|.$$
	A similar treatment, to that given in case ii), is given in cases iv),vi) and viii) ; and in all of them it can be obtained that
	$$ | \alpha (	f(\bar{x}_1+x) - f(\bar{x}_2+x)) | \leq \alpha |a| |\bar{x}_2 - \bar{x}_1|.$$
	For case iii) $f(\bar{x}_1+x) - f(\bar{x}_2+x)= b (\bar{x}_1 - \bar{x}_2)- 2 (a-b)$. Now, $a<b<0$ implies that 
	\begin{equation}\label{h3}
	b (\bar{x}_1 - \bar{x}_2)- 2 (a-b)	 \geq  b (\bar{x}_1 - \bar{x}_2 ).
	\end{equation}
	In this case $\bar{x}_1 - \bar{x}_2 \leq - 2$. Then
	\begin{equation} \label{h4}
	\begin{aligned}
	b\left(\bar{x}_{1}-\bar{x}_{2}\right)-2(a-b) & \leqslant b\left(\bar{x}_{1}-\bar{x}_{2}\right)+(a-b)\left(\bar{x}_{1}-\bar{x}_{2}\right) \\
	&=a\left(\bar{x}_{1}-\bar{x}_{2}\right).
	\end{aligned}
	\end{equation}
	From (\ref{h3}) and (\ref{h4}), we may conclude that
	$$ | \alpha (	f(\bar{x}_1+x) - f(\bar{x}_2+x)) | \leq \alpha |a| |\bar{x}_2 - \bar{x}_1|.$$
	In a similar way the latter estimate is also obtained for the case vii). \\
	Finally, taking into consideration the estimates obtained in each case, we can conclude that $H$ satisfies
	\begin{equation*}
	\norm{ H(t, \mu, \mathbf{z_2})-H(t, \mu,\mathbf{z_1})}\leq L\norm{ \mathbf{z_2}-\mathbf{z_1}},
	\end{equation*}
	with $\mathrm{L}= \alpha \max ( |a|, |b|)= \alpha |a|$.
\end{proof}

\bibliographystyle{plain} 
\bibliography{references} 
\end{document}